\documentclass[12pt,a4paper]{article}

\usepackage{fullpage,amsmath,amssymb,amsfonts,epsfig,setspace,apacite}
\psfigdriver{dvips}

\begin{document}

\title{\begin{bfseries}{The Tangled nature model with inheritance and constraint: Evolutionary ecology restricted by a conserved resource}\end{bfseries}}

\author{Simon Laird\footnotemark[1], Henrik Jeldtoft Jensen\footnotemark[1] \footnotemark[2]}

\maketitle
%\linenumbers

\begin{abstract}
We study a version of the Tangled Nature model of evolutionary ecology redefined in a phenotype space where mutants have properties correlated to their parents. The model has individual-based dynamics whilst incorporating species scale competitive constraints and a system scale resource constraint. Multiple species arise that coexist in a species interaction network with evolving global properties. Both the mean interaction strength and the network connectance increase relative to the null system as mutualism becomes more extensive. From a study of the dependence of average degree on the resource level we extract the diversity-connectance relationship which conforms to the hyperbolic form seen in field data. This is adjudged to arise as a consequence of the evolutionary pressure to achieve positive interactions. The network degree distributions conform more strongly to exponential than to the null binomial distributions in all cases. This effect is believed to be caused by correlations in the reproductive process. We also study how resource availability influences the phenotypical lifetime distribution which is approximately of power law form. We observe that the mean lifetime is inversely related to the resource level.
\end{abstract}

\section{Introduction} \label{sec.introduction}

\fnsymbol{footnote}
\footnotetext[1]{Department of Mathematics, Imperial College London,  South Kensington campus,  London SW7 2AZ, U.K.}
\footnotetext[2]{Author for correspondence (h.jensen@imperial.ac.uk)}

A functioning ecosystem relies on various environmental resources to sustain its component species with some of these resources being universally required for survival. The limitation of supply rate or overall quantity of such a resource, acts as a restriction to the development of an ecosystem and systematic properties are expected to depend on the level at which this occurs \shortcite{loeu03:nutr}. It is known from field data that the diversity of species changes across energy and water gradients such as is found in latitudinal variations \shortcite{hawk03:ener} \shortcite{bonn04:stru}. In general, heterogeneous environments will support greater species numbers at higher resource availabilities with the relationship taking a monotonically increasing form \shortcite{curr91:ener} \shortcite{waid99:rela} \shortcite{bonn04:stru}. This difference in diversity and the knock-on effect of elevated competition could conceivably lead to a change in the topology and dynamics of the species interaction network. These networks and their associated interaction matrices have been studied and are considered to be significant in adjudging the stability and permanence of ecosystems \shortcite{may74:stab} \shortcite{treg79:comp}. 

There has been a great deal of research investigating the short term effects of resource variation on ecosystems \shortcite{tilm:reso} \shortcite{hulo00:func}, but very little on the impact of resource levels over evolutionary timescales. Also, recent work has focused on the dynamical coevolution of species populations \shortcite{toki03:emer} \shortcite{dros04:impa} \shortcite{mcka04:evol} \shortcite{copp04:exti} but these approaches use continuum descriptions whilst evolution occurs via dynamics associated with discrete individuals. To address these issues accordingly we present a model of evolutionary ecology that functions at the level of the individual whilst incorporating species scale competitive constraints and a system scale resource constraint. 

Our work follows from the Tangled Nature model of evolutionary 
ecology \shortcite{chri02:tang} \shortcite{hall02:time} \shortcite{diCo03:tang} \shortcite{ande05:netw}, which demonstrated species 
formation as a result of individual-based dynamics. These species were then studied in terms of their 
interaction networks and other more dynamical features such as extinctions, lifetimes and stabilities. 
Recently, the model was simplified by restricting the reproductive process to non-overlapping 
generations \shortcite{rikv03:punc} \shortcite{zia03:fluc}. This allowed both longer timescales to be simulated 
and a deeper analytical treatment to be made available. Modifications made to the description of individuals 
have now allowed us to include a gradual species evolution within a significantly larger {\it{phenotype}} space. 
Correlations introduced into the structure of this space mean that inheritance from parent to mutated offspring 
is available and also mathematically quantifiable which was not the case in the previous model. Attempts have been 
made to impose correlations onto the hypercubic genotype structure of the earlier model \shortcite{sevi05:effe} 
but with only limited effect. Here we use a different procedure to construct a correlated phenotype
space which allows us to consider much larger spaces. We find that correlations in these huge phenotype
spaces have a significant effect.

Our paper is arranged as follows. In the first part of section \ref{sec.model}, we describe how the system members are represented in terms of a hypothetical phenotype space. This representation is used to provide a fully determined, correlated interaction-space. As the processes involved in this are somewhat laborious we have consigned the wider details to the appendix, whilst giving an overview in the main text. In the second part of section \ref{sec.model}, we describe the update rules for the system and how the reproductive probabilities are determined from the species interaction network. We approach the dynamical evolution in a stochastic discrete manner as we see it as appropriate for a discrete-entity system such as this. Statistical fluctuations are important as population numbers may become low and coupled dynamics across an interaction network could be heavily dependent upon them. 

Results regarding the issues of diversity, lifetime distributions and network properties are presented in section \ref{sec.results}, and a discussion of these along with future directions are set out in section \ref{sec.discussion}.

\section{Model} \label{sec.model}
\subsection{Interactions} \label{sec.interactions}

Individuals are represented by vectors, $\bf{T}^{\alpha}=(\it{T}^{\alpha}_{1},\it{T}^{\alpha}_{2},...,\it{T}^{\alpha}_{L})$, in a phenotype space of dimension $L=16$, with specific individuals denoted by Greek lettering $\alpha,\beta,...=1,2,...N(t),$ where $N(t)$ is the population at time, $t$. Each trait $\it{T}^{\alpha}_{i}$ may take an integer coordinate in the range [0,99999] that is periodically bounded allowing the points ${\it{T}^{\alpha}_{i}}=0$ and ${\it{T}^{\alpha}_{i}}=99999$ to be contiguous. The phenotype coordinates themselves are arbitrary and are not intended to represent any kind of quantitative scale so the periodic boundary is introduced to maintain their arbitrariness. A large integer range is used to emulate the continuous nature of many real organism traits and as a result the extant phenotypes are capable of gradual adaption during the evolutionary process.

The co-evolution of the extant phenotypes is primarily controlled by the interactions they have with one another. These are represented by ${\rm{J}}^{\alpha\beta}={\rm{J}}(\bf{T}^{\alpha},\bf{T}^{\beta})$, the interaction strength effected upon $\alpha$ by $\beta$, which is independent of ${\rm{J}}(\bf{T}^{\beta},\bf{T}^{\alpha})$. As the phenotype space is a closed set of ${100000}^{16}={10}^{80}$ possible states the interaction space is a closed set of ${10}^{80}{\times}{10}^{80}={10}^{160}$ possible pairwise interactions that exist {\emph{in potentia}}. This set of all possible phenotype interactions is constructed with the property that only a subset of them are non-zero quantities. This proportion or {\emph{connectance}}, ${\theta}_{0}$ is defined at the outset and represents the fact that a specified phenotype will interact only with a subset of all other conceivable phenotypes. The resulting set of non-zero interactions are assigned strength values that are normally distributed.

A system created with a random group of phenotypes will, on average, conform to the above properties (ie. connectance, ${\theta}={\theta}_{0}$, plus interaction strengths that are normally distributed). But, once the evolutionary process is underway the phenotypes existing at later times may have interaction properties that deviate from those of the random null system. The fact that we pre-define the properties of the closed interaction space means that the evolving system properties can be compared quantitatively to the null system.

The phenotype space is constructed in a manner that gives a quantifiable correlation, ${\bf{C}^{\alpha\beta}}$ between separate points. This correlation measure decays approximately exponentially with separation in the phenotype space and has an effect on the inheritance of both the interaction set distribution (the subset of non-zero interactions) and the strength measures themselves. The consequence is that any mutated offspring will have a similar interaction set and strengths to those of the parent, with the level of similarity being dependent upon the mutation length.   

The requirement of a large, correlated, pre-defined interaction space means using a deterministic procedure for producing the interaction strengths. Any interaction between defined phenotypes must be time-invariant so as there are many more possible interactions than can be stored computationally we used a deterministic method for acquiring the values from any given pair of phenotype vectors. The exact method, whilst not difficult, is somewhat convoluted so is explained separately in the appendix along with the details of how the correlations are imposed.

Intra-specific competition is deemed a necessary part of the dynamics to allow diversity and is incorporated by setting all self-interactions to a negative value that is constant and independent of the phenotype vector. This property extends via the imposed correlations to phenotypes of similar but distinct vectors thus allowing for the localised but distributed nature of a species description. 

\subsection{Dynamical update} \label{sec.dynup}

We initiate the system with a set of individuals which are assigned random phenotype vectors. At each subsequent timestep an individual is selected for annihilation with probability $P_{kill}=0.2$ whereby it is removed from the system and the single resource unit associated with it is returned to the resource bath, $R(t)$. During the same timestep, another individual is randomly selected to reproduce with probability $P_{repro}$. This value is determined via use of the weight function,

\begin{equation}\it{H}(\bf{T}^{\alpha},\it{t})={a_{1}}{\frac{\sum^{N(t)}_{\beta=1}{\rm{J}}^{\alpha\beta}}{\sum^{N(t)}_{\beta=1}\bf{C}^{\alpha\beta}}}-{a_{2}}{\sum^{N(t)}_{\beta=1}\bf{C}^{\alpha\beta}}-{a_{3}}{\frac{N(t)}{R(t)}}.\label{eq.hamiltonian}
\end{equation}

This sum is monotonically mapped to the interval [0,1], appropriate for a probability measure, through use of the function,

\begin{equation}P_{repro}=\frac{exp[\it{H}(\bf{T}^{\alpha},\it{t})]}{1+exp[\it{H}(\bf{T}^{\alpha},\it{t})]}.\label{eq.compfermi}
\end{equation}

The first term in Eq.(\ref{eq.hamiltonian}), represents the summed effect of interactions with other system members. Although it is made over all individuals at time, $t$, many of the contributions will be zero as the two phenotypes in question may not interact. This is an inherent feature of the complete interaction set which permits only a prescribed proportion, $\theta_{0}$, of all possible phenotype couplings to exist (as evolution occurs though, the connectance of the extant system, $\theta$ may deviate from, $\theta_{0}$).

The denominator represents the number of individuals that can be described as belonging to the same species group as $\bf{T}^{\alpha}$. This damping effect is used to prevent divergence, but can be seen as realistic if we consider the sharing of the interaction effects amongst members of the same species. An example would be the predatory effect of a single lion amongst a herd of wildebeest. As the lion feeds up to a maximum rate the effect of the lion on an individual in a small herd is more pronounced than the effect in a large herd. Essentially, the probability of a specific lion consuming a specific wildebeest decreases in relation to the size of the herd. This damping effect is by no means ubiquitous but it is nevertheless a feature of many interactions. 

Phenotypes that are in close proximity in the phenotype space are distinct but correlated in a manner that can be described as a species set, so this must be accounted for here. This is achieved by using ${\bf{C}}^{\alpha\beta}\in[0,1]$, which is a measure of correlation between $\alpha$ and $\beta$. A more detailed explanation is given in the appendix.

The second term of Eq.(\ref{eq.hamiltonian}), represents conspecific competition so utilises the same correlation measure as the denominator in the first term. This negative contribution is essential to allow diversity in the system and generalises the multitude of factors other than the conserved resource that members of the same species compete over.

The third term of Eq.(\ref{eq.hamiltonian}), represents the general system competition for the conserved resource. It represents the number of individuals, $N(t)$, that are competing per unit of available resource, $R(t)$. As the condition for conservation, $R(t)+N(t)=constant$ must hold, the numerator and denominator are dependent. The resource range investigated here spans from 1000 to 30000 units.

The parameters, $a_{1}=0.5$, $a_{2}=0.01$, $a_{3}=0.2$ are the selection, conspecific competition and resource competition parameters respectively. These values were subjectively selected to allow significant diversities to be achieved whilst incurring the dynamics to be strongly dependent upon inter-specific interactions. 

 Upon successful reproduction, the individual produces a single offspring that assumes one unit of resource from the resource bath. As the bath depletes the probability of reproduction for any phenotype is reduced due to the third term in Eq.(\ref{eq.hamiltonian}). The offspring phenotype is identical to that of the parent (clonal) unless, with probability $P_{mut}=0.0002$, a mutation occurs. This is effected by shifting a randomly selected trait value by a random amount ${\psi}$, that is gaussian distributed with $\mu=0$ and ${\sigma}=\xi$, where $\xi=250$ is the correlation length of the phenotype space. Any mutated offspring will have an interaction set that is similar but not identical to its parent in accordance with the correlation. We have elected to use a low mutation rate with a relatively high mutation distance to allow the evolution to occur efficiently enough for tractable simulation times.

\section{Results}\label{sec.results}
\subsection{Diversity}\label{sec.diversity}
By initialising the system with ${\theta}_{0}=0$ we can generate a neutral evolution in which the population increases whilst diffusing uniformly through the phenotype space. The intra-specific competition term causes this diffusion by forcing the system members to be as little correlated as possible. As a result, the diversity grows to large values whilst localised phenotype populations remain low. The incorporation of a non-zero ${\theta}_{0}$ breaks this symmetry allowing phenotypes to counteract the competitive constraint with positive interactions and so accumulate localised populations. These phenotypes are distributed as highly populated single sites surrounded by a sparse cloud of mutants that derive primarily from the central 'wild type'.
\begin{figure}
[ptb]
\begin{center}
\epsfig{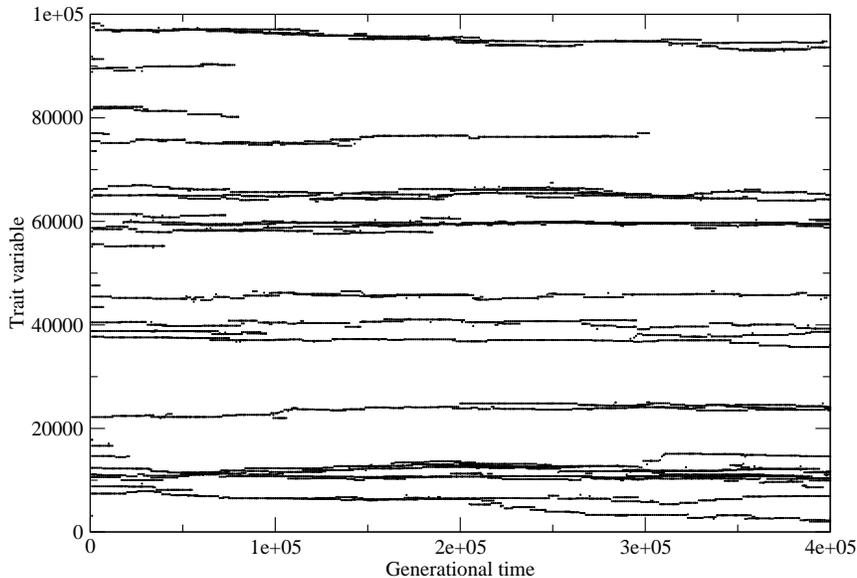}
\caption{\small{An occupation plot of a single run for a system with R=10000. For each timeslice a point appears where a phenotype is in existence but as the full space is in 16 dimensions a projection onto a single trait is used.}}\label{fig.occplot}
\end{center}
\end{figure}

Fig.(\ref{fig.occplot}), shows a section of the time evolution of the extant species in a single run of one million generations. The visual representation of this is as a projection of the populated points of the 16 dimensional phenotype space onto a single trait. It is clear to see that the evolutionary process creates a system that is far from diffuse with a small set of phenotypes interacting in a manner that precludes easy invasion by mutants. Of course, there are successful invasions that amount to gradual evolution of a species, or even speciations, but the relative permanance of species is seen as significant. This is because a new mutant phenotype will have an advantage over the parent due to the relative weakness of its intra-specific competition term, so a continual invasion of species could easily be expected. The phenotype distribution localises at points rather than following a diffusive process and does so to quite an extreme. There is nothing to prevent the diversity from expanding with species achieving smaller populations but this state might struggle to persist. It is likely that the stochasticity of the dynamics would incur a greater extinction rate for species of such sparse numbers thus reducing the diversity. This is one reason proposed to explain why productivity-diversity relationships having increasing functional forms at low productivity ranges \shortcite{pres62:cano} \shortcite{abra95:mono}.

\begin{figure}
[ptb]
\begin{center}
\epsfig{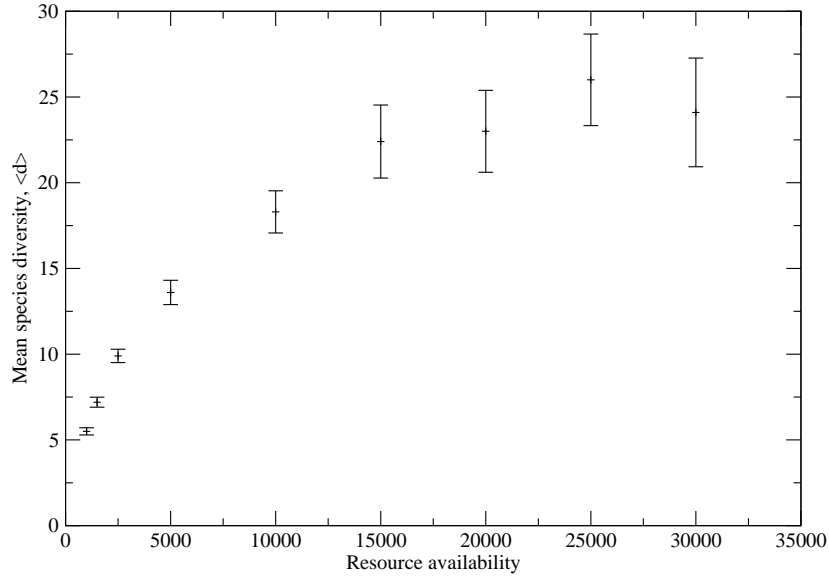}
\caption{\small{Plot of mean species diversity in relation to resource availability. Error bars represent the standard error.}}\label{fig.div}
\end{center}
\end{figure}

The diversity varies considerably both in time and across actualisations. This is most apparent for higher resource levels where the standard deviations of the diversity become comparable to the means. Regardless of this spread, for the range investigated here the mean species diversity increases with respect to total resource availability in a monotonic fashion, Fig.(\ref{fig.div}). 
This relationship has been produced in a species level trophic network model \shortcite{mcka04:evol} and is empirically found in large scale systems with heterogeneous environments \shortcite{curr91:ener} \shortcite{waid99:rela} \shortcite{bonn04:stru}. Although unimodal relationships are expected for localised ecosystems where diversity is more dependent upon fewer limiting factors this model incorporates aspects that are best compared to heterogeneous larger scale systems. At this ecological scale a monotonically increasing diversity can in some cases be ascribed to the effects of intra-specific density dependence \shortcite{abra83:theo} \shortcite{vanc84:inte} \shortcite{abra95:mono}. Resource increases allow species to grow in population but other factors more unique to that species' niche restrict this growth prior to the resource depletion becoming a limiting factor. The consequence is that resource is more freely available for species holding dissimilar niches that would be excluded at low resource due to their inferior ability to procure it. This model represents such systems as the intra-specific competition is the dominant restrictive term in Eq.(\ref{eq.hamiltonian}), for a high species population.

\subsection{Lifetimes and extinction rates}\label{sec.lifetimes}
\begin{figure}
[ptb]
\begin{center}
\epsfig{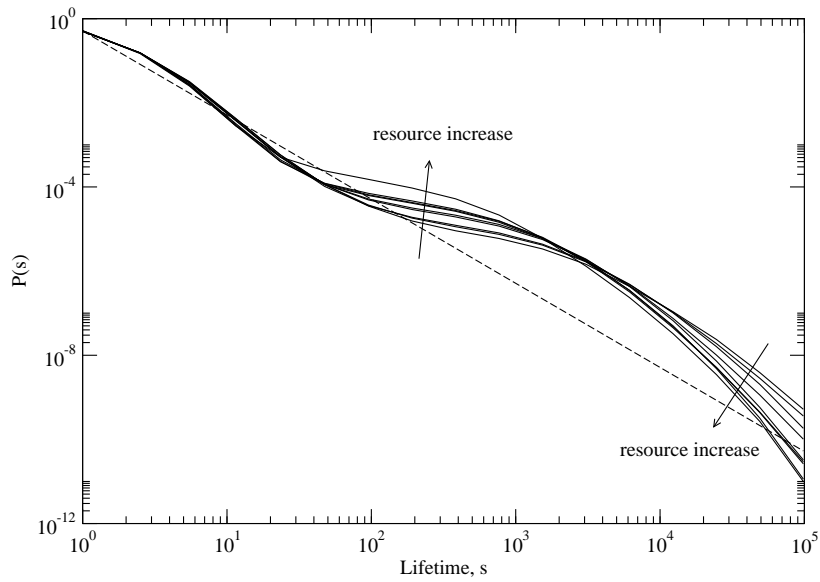}
\caption{\small{Plot of the lifetime distributions over all resource availability levels. In the approximate range $s=[10^2,10^3]$ the probability of a given lifetime increases with resource availability, whilst the reverse is true for the approximate range $s=[10^4,10^5]$. This indicates longer lived species at lower resources. The dotted line represents the often proposed power law relationship with exponent, $\alpha=2$.}}\label{fig.lifetimes}
\end{center}
\end{figure}

Statistical analyses of fossil record data have often alluded to power law forms, $P(s)\,\,{\sim}\,\,s^{-\alpha}$, in the distributions of the various quantities involved with species extinctions. When $s$ represents species lifetimes or extinction event sizes an exponent of ${\alpha}\,\,{\simeq}\,\,{2}$ has been suggested but the analyses have been criticised and power law forms are not readily accepted \shortcite{newm99:mode} \shortcite{dros01:biol}. The lifetime distributions produced in this model are clearly not of this form although they do loosely follow a power law with a comparable exponent Fig.(\ref{fig.lifetimes}).
 Other models have displayed similar off-power law or non-power law behaviour \shortcite{ecolab:ws} \shortcite{chri02:tang} \shortcite{chow03:unif} \shortcite{chow03:food} (and also \shortcite{chow03:sole} in reference to \shortcite{sole96:exti}) giving credibility to the notion that these distributions may not actually be scale free in the real system. The power law form seen in real macroecological data is tenuous as a result of the difficulty in appropriating quality data sets, but it may be the case that the actual distributions are really deviations from this form. It is hard to draw conclusions from models such as these as they are obviously simplifications of a highly complex, spatially-extended, multiscaled system but there is no {\it{a priori}} reason why power law behaviour should be expected anyhow. 

\begin{figure}
[ptb]
\begin{center}
\epsfig{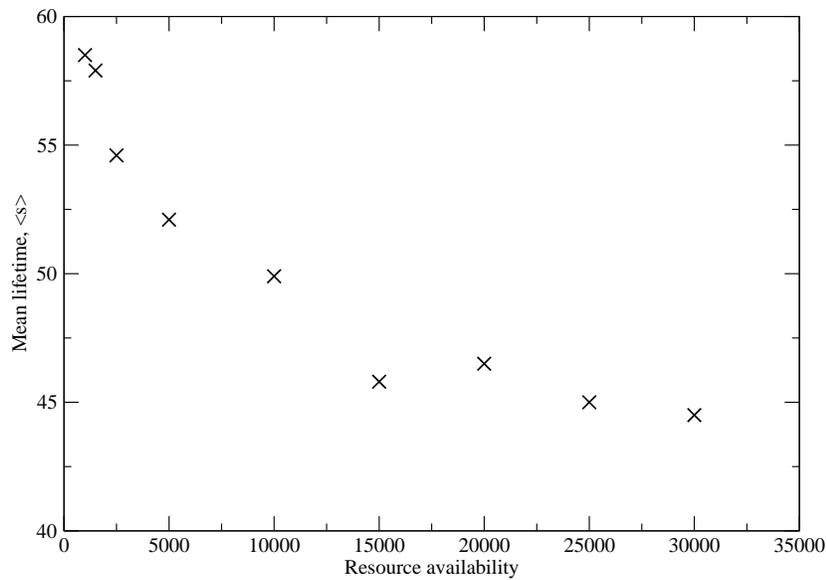}
\caption{\small{Plot of mean lifetime against resource availability.}}\label{fig.meanlife}
\end{center}
\end{figure}

\begin{figure}
[ptb]
\begin{center}
\epsfig{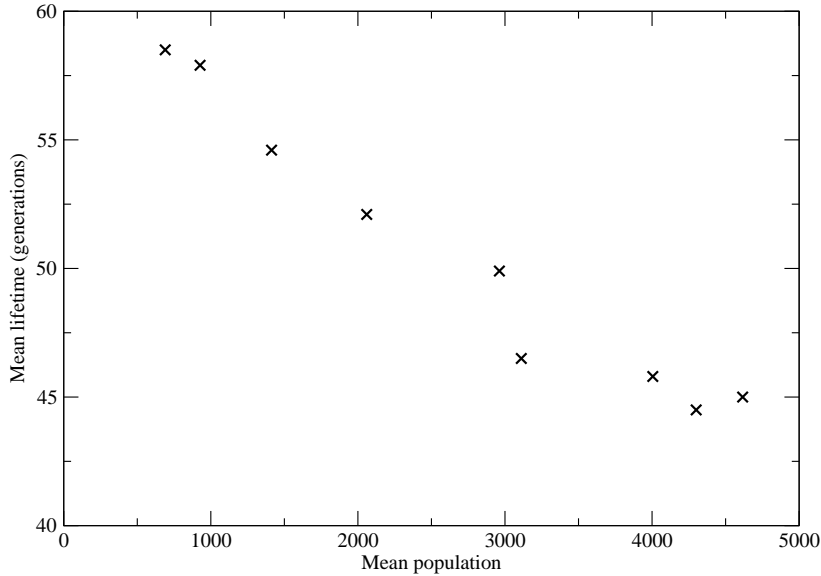}
\caption{\small{Plot of mean lifetime against mean population for each resource level.}}\label{fig.lifepop}
\end{center}
\end{figure}

Although the structural form of the distributions for different resource levels are similar there is a gradual decrease in the level of species permanence as resource increases. This effect can be seen in Fig.(\ref{fig.lifetimes}), as an increase in mid logarithmic-range lifetime probability accompanied by a sharper declining tail. It is a minimal but consistent trend, as can be observed in the relationship between mean lifetime and resource level in Fig.(\ref{fig.meanlife}). We can attribute this behaviour to the difference in mutant offspring production rates incurred at different resource levels. As the mean population increases with increasing resource availability there is a higher magnitude of successful reproduction events in each generational time step. So, given a constant mutation probability, the number of mutated offspring produced in this period increases in proportion to the population. The permanence of individual species and the stability of the interaction network are affected by this elevated mutant production rate leading to a greater rate of extinction and speciation. 
If we plot the mean lifetime of a given resource level against the mean population of that level Fig.(\ref{fig.lifepop}) a linear dependence can be seen that supports the explanation above. To justify this further we have run simulations with variable mutation probabilities that force the absolute mutant production rate to remain constant. The lifetime distributions produced from these runs show no dependence on resource availability which is consistent with our reasoning.

\subsection{Network properties}\label{sec.network}
An important feature of a coevolving ecosystem is the species interaction network, particularly as it is thought to have implications for the stability of the ecosystem and the permanence of its members. Low diameter networks with scale invariant degree distributions have often been expected as many abiotic networks display such properties and there are theoretical reasons why they may augment stability \shortcite{albe02:stat} \shortcite{doro02:evol}. In reality, field data has shown the existence of a variety of distributions including power law, truncated power law and exponential \shortcite{dunn02:food} \shortcite{jord03:inva}, but these still differ from the binomial distributions found in finite random networks. 

The degree distributions produced by our model take an exponential form which is similar, at low degrees, to the binomial distribution of the same connectance. The difference becomes apparent at higher degrees where the binomial decays much more rapidly than the exponential.

\begin{figure}
[ptb]
\begin{center}
\epsfig{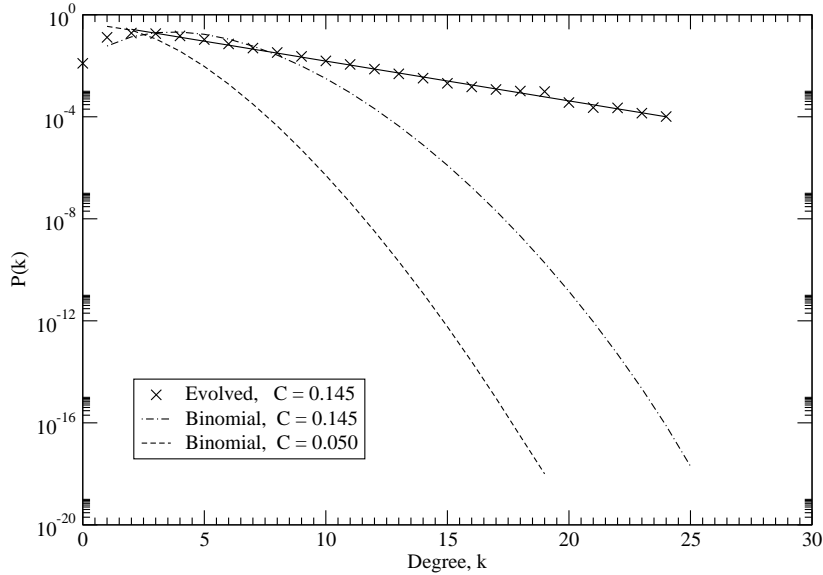}
\caption{\small{Plot of degree distribution for the 30000 unit resource system. Compared are the binomial distribution of the same connectance, C, and the binomial distribution expected for the null case with the space connectance $C={\theta}_o=0.05$}}\label{fig.degree}
\end{center}
\end{figure}

Fig.(\ref{fig.degree}), shows the degree distribution for the realisations with diversity of 29 ($R=30000$) along with a binomial distributed plot using the same connectance. It is clear that the evolved system has a distribution with a much longer tail than the binomial. 

The deviation of the degree distribution away from the binomial can be attributed to the inheritance-based development of the network structure. Simulating the evolution using uncorrelated mutation leads to binomial distributions, albeit with enhanced connectance. The incorporation of correlated inheritance has the effect of producing a longer tailed distribution that conforms closely to an exponential.

\begin{figure}
[ptb]
\begin{center}
\epsfig{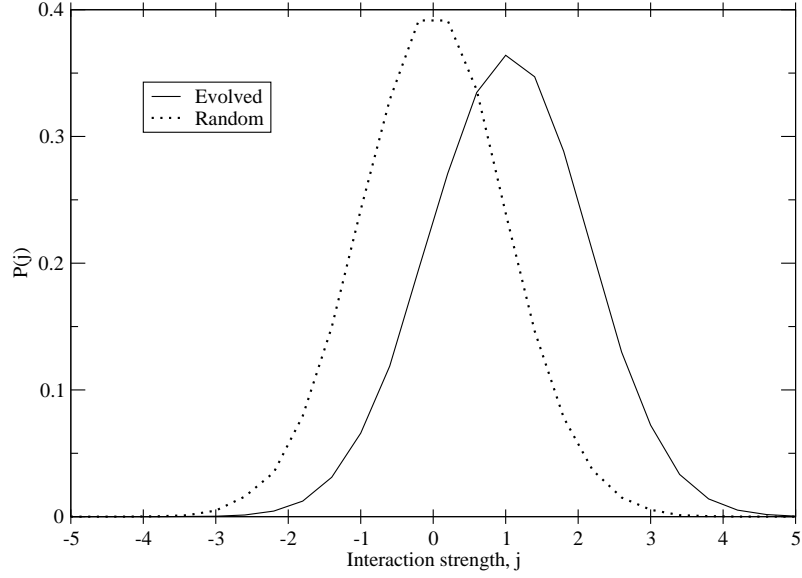}
\caption{\small{Plot of the interaction strength distribution for the 30000 unit resource system. The evolved distribution deviates from the null normal towards positive strengths.}}\label{fig.istrength}
\end{center}
\end{figure}

\begin{figure}
[ptb]
\begin{center}
\epsfig{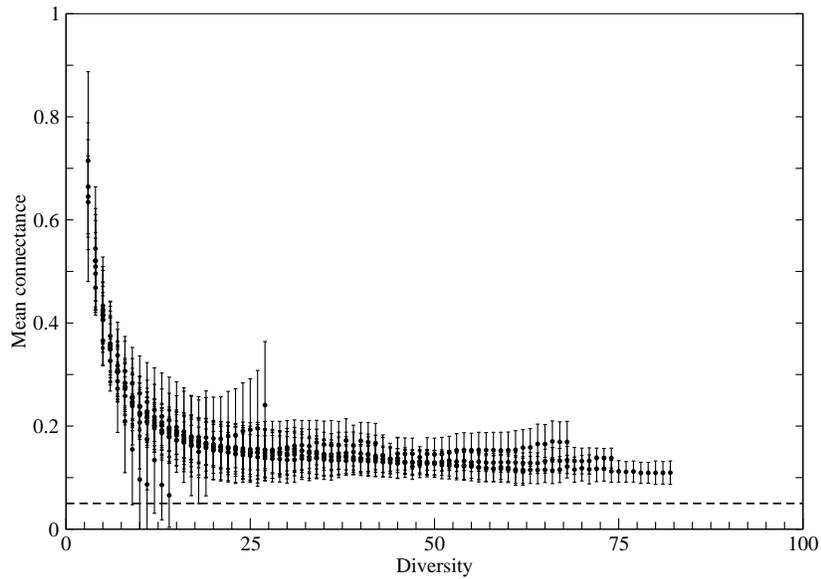}
\caption{\small{Plot of mean connectance, $<{\theta}>$ against diversity. Each resource level is treated separately so multiple data points often occur for the same diversity value. Error bars represent the standard error. The lower dotted line marks the null system connectance, ${\theta}_{0}=0.05$}}\label{fig.connectance}
\end{center}
\end{figure}

A notable effect of the evolution is the dramatic increase in the mean degree compared to that of a random species set of the same diversity. There are a great deal more connections between the extant species than is found in the null case. This is demonstrated in Fig.(\ref{fig.degree}), where the evolved distribution can be compared to the binomial distribution that would be found for a random set of 29 species. The connectance values for all diversities are very much higher than general connectance of the interaction space ${\theta}_{0}$ (see Fig.(\ref{fig.connectance})) and they easily exceed the percolation thresholds that would be assigned to networks of these node/diversity numbers \shortcite{albe02:stat}. It is unclear whether greater stability is achieved through either lower than threshold connectance values or higher values \shortcite{may74:stab} \shortcite{mcca00:dive} \shortcite{hayd00:maxi} \shortcite{dunn02:food}, but here the system naturally evolves towards the latter. It appears that the system has a strong tendency to evolve towards configurations that are highly mutualistic by assuming greater numbers of positive interactions. This is highlighted by the interaction strength distribution that has clearly shifted from the null system normal with zero mean to a Gaussian form with a positive mean, Fig.(\ref{fig.istrength}). 
As we increase resource, and therefore the diversity, the mean connectance  decreases as the system becomes progressively less likely to achieve large deviations from the null system expectation values, Fig.(\ref{fig.connectance}). In the limit of large diversity the connectance would of course conform to that of the interaction space, $\theta_{0}$.

\section{Discussion}\label{sec.discussion}

The concepts used in this model have been developed from original ideas presented in the Tangled Nature model  \shortcite{chri02:tang} \shortcite{hall02:time} \shortcite{diCo03:tang} \shortcite{ande05:netw}, where species interaction networks were constructed from dynamics instigated at the level of individuals as opposed to higher level groupings such as species. Also, the use of a predefined, closed set of possible interactions between species allowed statistical comparisons to be made between evolved systems and null species sets. These principles have been incorporated in the new model along with modifications to improve the reality of the dynamics.

The main modification is in the creation of the phenotype space. This has permitted a much larger set of distinct entity types to be available, $10^{80}$ as opposed to $10^{20}$, and allowed us to introduce mathematically quantifiable correlations in the inheritance of interaction properties between parent and mutated offspring. Along with a more continuous phenotype description, these attributes make the model more representative of a real biological system. 

The new model includes a generalised niche property in the form of an intra-specific competition term in the weight function. It was realised that this was necessary to allow a diversity to flourish without competitive exclusion reducing the system to a pair of highly mutualistic species. We make no attempt to describe any niche properties nor numbers of niches but the inclusion of this concept can be justified as a simplistic representation of a heterogeneous environment. 

With the above qualities in effect the primary aim of the research here is to investigate how the constraint of a conserved resource effects the dynamics and interaction properties of the species evolution. The mean species diversity exhibits a monotonic increase with respect to system resource availability which is expected for a real system with a heterogeneous environment. This increase is also accompanied by a greater level of fluctuation which is consistent with a reduction in the degree of species permanence, as implicated by the lifetime distributions. Although the effect is small due to the inclusion of all mutant phenotypes in the distribution statistics, there is a clear decrease in mean lifetimes as we increase the resource level. To ensure this decline is not simply an artefact of the higher diversity systems producing more short-lived mutants, which are included in the statistics here, we repeated the analyses omitting contributions below various threshold lifetime values. In each case the decline was observed. We can infer from this that the decrease in mean lifetime is applicable to the more permanent members which is more representative of a real system. Species considered in the collection of field data would not be random mutants, but recognisable community members. If there exists a higher species turnover rate at higher resource levels then it is conceivable that the evolutionary rate is enhanced as well. There is evidence of a positive relationship between evolutionary rate and energy supply but the mechanisms suggested for its existence are not related to the concepts here \shortcite{pawa05:geog}. It has recently been discovered though that higher diversity systems may incur a greater turnover rate of species and so a smaller duration probability \shortcite{emer05:spec}. Our work certainly points to the same conclusion. 

The realised species interaction networks universally conform to exponential degree distributions which have also been reported in some real ecosystems. The long tails of these exponential distributions indicate the existence of small numbers of massively connected species which is highly unlikely to occur in the random binomial distributions. Simulations performed with random mutations exhibit binomial degree distributions, which implies that the correlations involved in inheritance are at the root of the exponential degree distributions. It is conceivable that the non-binomial distributions found in real ecological networks appear for similar reasons. The component species of these networks, whilst not necessarily speciations of one and other, are likely to have traits or properties that could be considered as correlated. Invasive forms of community assembly would lead to systems with lower levels of correlation and so may exhibit less pronounced deviations from the binomial as a result.

%It has been suggested that there is a correlation between link density and abundance of a species \shortcite{vazq04:asym} \shortcite{vazq05:spec} with implications as to the form of the species abundance distribution. There is some uncertainty in this type of distribution here but there is evidence of an underlying pattern which is currently being investigated. 

The high connectivity of the {\it{hub}} species contributes to the overall connectance of the network which evolves to statistically large values. It is unclear whether this level of connectance affects the stability and permanence of the networks as the system is continually attracted to achieving connections. It is interesting though that the mean connectance drops to a near constant level as higher resources/diversities are approached. This type of behaviour has been alluded to in field studies \shortcite{mart92:cons} \shortcite{dros02:mode} \shortcite{mont03:topo} with a power law functional form suggested for the connectance-diversity relationship. A hyperbolic form has been suggested \shortcite{mcka00:mean} for reasons of stability with deviations from the inverse power law being attributed to fluctuations via species invasions. Our model, due its inclusion of mutualistic interactions, gives an alternative explanation to this phenomenon. Our systems evolve in a manner that intrinsically construct networks with greater numbers of interactions. As diversities grow, the probability of achieving such high connectances diminishes leading to the observed relationship between the two properties.

Making comparisons between food webs and the more encompassing species interaction networks of our model is problematic though. The mutualistic interaction is an important feature here and is vital to the proposition that connectance is an evolutionarily favoured global property. Most field data is predator-prey based and so precludes those types of interaction. The studied networks may well be embedded within larger species interaction networks that involve mutualists and commensalists though so our reasoning still has validity.

\section{Appendix}

In order to create a predetermined and correlated interaction space the following process is used. We consider the interaction strength of $\alpha$ due to $\beta$, ${\rm{J}}^{\alpha\beta}$, to be determined additively through contributions associated with each of the $L$ traits,

\begin{equation}{\rm{J}}^{\alpha\beta}=\frac{1}{L^{\frac{1}{2}}}\sum^{L}_{i=1}\it{J}^{\alpha\beta}_{i}.\label{eq.jsum}
\end{equation}

The contributions $\it{J}^{\alpha\beta}_{i}$ will take values that are normally distributed so the prefactor $\frac{1}{L^{\frac{1}{2}}}$ is used to ensure the final strengths are similarly distributed.

To acquire these separate contributions we first couple each trait $\it{T}^{\alpha}_{i}$ to all others, from both $\alpha$ and $\beta$ sets, to provide $L$ {\it{modified}} trait variables, $\it{T}^{\alpha\beta}_{i}$. This is achieved through the sum,

\begin{equation}\it{T}^{\alpha\beta}_{i}=\sum^{L}_{j=1}{b_{ij}}\it{T}^{\alpha}_{j}+\sum^{L}_{k=1}{c_{ik}}\it{T}^{\beta}_{k}\,\,\,\,\,\,\, mod(100000),\label{eq.traitsum}
\end{equation}

where ${b_{ij}},{c_{ik}}\in{\{-1,1\}}$ are independent, randomly assigned with equal probability and held constant throughout. This coupling format allows any shift (mutation) in a single trait variable to lead to a shift of the same magnitude in all modified trait variables. The coefficients, ${b_{ij}},{c_{ik}}$ ensure that the resultant shifts in the modified trait variables are different for each initially mutated trait. The intention here is to map these modified trait variables to values that will represent the separate contributions to the interaction strength of Eq.(\ref{eq.jsum}). As will be demonstrated, the couplings assist in producing a quantifiable correlation, but they may also be justified as representing real interactions. The efficacy of a trait is dependent upon both its own, and the other interacting species, trait sets. For example, in pursuit of prey, the cheetah's high speed would facilitate its success but be dependent upon its stealthiness and the prey's capability of flight. The traits are interdependent and so lead to an interaction strength (or similarly, fitness) in a coupled manner.

The separate contributions to the interaction strength of Eq.(\ref{eq.jsum}), are now determined by mapping each modified trait variable to a strength value,

%\begin{equation}
%{\it{J}}^{\alpha\beta}_{i}={\cal{L}}({\it{T}}^{\alpha\beta}_{i}).
%\end{equation}

\begin{equation}
{\cal{L}}_{i}{\colon}{\it{T}}^{\alpha\beta}_{i}{\mapsto}{\it{J}}^{\alpha\beta}_{i},i\,{\in}\,[1,16].
\end{equation}

%%%%%%%%%%%%%%%%%%%%%%%%%%%%
% start of additional part %
%%%%%%%%%%%%%%%%%%%%%%%%%%%%

%$\it{J}^{\alpha\beta}_{i}$ is then determined by mapping $\it{T}^{\alpha\beta}_{i}$ to a trait-specific, normally distributed first order autoregression series ${\cal{L}}_{i}{\colon}{\it{T}}^{\alpha\beta}_{i}{\mapsto}{\it{J}}^{\alpha\beta}_{i},i\,{\in}\,[1,16]$. Here ${\cal{L}}_{i}$ refers to an independent series for each modified trait, i. These series are created using the stochastic iterative Markov process,

%\begin{equation}
%x_{n+1}={\rho}x_{n}+{\phi}
%\end{equation}

%where $\phi$ is a normally distributed random variable, $\rho$ determines the correlation length of the series, $\xi=250$, and $n{\in}[0,99999]$. This form of series has the overall property that its values conform to a normal distribution (once normalised) so providing trait-specific interaction strengths $\it{J}^{\alpha\beta}_{i}$ that distributed in the same manner.

\begin{figure}
[ptb]
\begin{center}
\epsfig{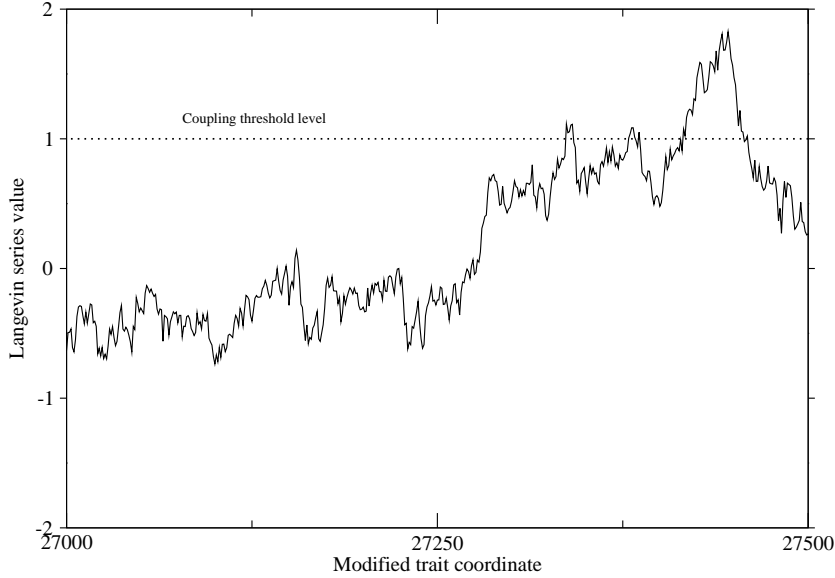}
\caption{\small{A section of a first order autoregression series. The values can be seen to have a degree of correlation. The series values are globally normally distributed so by use of the error function we can impose a threshold mark (straight line) which is a requirement for a coupling to be deemed to exist }}\label{fig.threshold}
\end{center}
\end{figure}

%%%%%%%%%%%%%%%%%%%%%%%%%%%%
% end of additional part   %
%%%%%%%%%%%%%%%%%%%%%%%%%%%%
 For this we use $L=16$ independent series indexed from 0 to 99999 to allow a one to one mapping between the modified trait variable and the strength measure (see Fig.(\ref{fig.threshold})). To incur correlations we use first-order auto-regression series \shortcite{hami94:time} that are created using the stochastic iterative Markov process,

\begin{equation}
x_{n+1}={\rho}x_{n}+{\phi}
\label{eq.markov}
\end{equation}

where $\phi$ is a normally distributed random variable, $\rho$ determines the correlation length of the series and $n{\in}[0,99999]$. This form of series has the overall property that its values conform to a normal distribution (once normalised) so providing trait-specific interaction strengths $\it{J}^{\alpha\beta}_{i}$ that are distributed in the same manner.

The correlation function of the series has an exponential decay,

\begin{equation}
{{\bf{C}}_{\cal{L}}=exp[-{\mid}n-m{\mid}/\xi]}, 
\end{equation}

where $n,m$ are the indices of the series values. The correlation length, ${\xi}=250$ is a metric of the index separation of the series values that is definable through the parameter, ${\rho}$ from Eq.(\ref{eq.markov}). With the use of full coupling of all traits to give a modified trait this function extends to separations in the phenotype space where the differences in trait coordinates sum to give an approximate two-point correlation function,

\begin{equation}{\bf{C}^{\alpha\beta}}\,\,{\simeq}\,\,{exp\,[\frac{-{\frac{1}{L}}{\sum^{L}_{i=1}}{\Delta}{T}^{\alpha\beta}_{i}}{\xi}]}\label{eq.correlation1}
\end{equation}

where,

\begin{equation}{\Delta}{T}^{\alpha\beta}_{i}={\mid}{b}_{i1}({T}^{\alpha}_{1}-{T}^{\beta}_{1})+{b}_{i2}({T}^{\alpha}_{2}-{T}^{\beta}_{2})+...+{b}_{iL}({T}^{\alpha}_{L}-{T}^{\beta}_{L}){\mid},\label{eq.correlation2}
\end{equation}

and ${b}_{ij}$ are the parameters used in Eq.(\ref{eq.traitsum}).

If a mutation is imposed by the shift of a single trait variable then the numerator of Eq.(\ref{eq.correlation1}) simply reduces to the length of that shift. Thus a mutated phenotype has a quantifiable correlation with the parent, with an exponential decay determined by the size of the trait shift. This is where the trait couplings of Eq.(\ref{eq.traitsum}) facilitate the quantification of the correlation.

In a real ecological system, not all phenotypes interact directly with one another so only a certain proportion of the set of all interactions is allowed in the model. This is achieved by using the same process used to determine the interaction strengths except with some important modifications. Using a further set of autoregression series the process is repeated and the interaction is deemed to exist if, and only if, the two phenotypes have mapped values above a certain threshold, Fig.(\ref{fig.threshold}). As these values are normally distributed the error function can be used to determine this threshold so as to set the connectance of the interaction space ${\theta}_0$. For a particular phenotype, the proportion of all possible interactions it has across the phenotype space will be equal to this value, barring statistical fluctuation. The distribution of this set throughout the space is quite diverse with patches of varying size due to the random but correlated nature of the autoregression series and the interdependency of the traits.

Phenotypic self-interaction is excluded in this model (intra-specific competition is considered separately Eq.(\ref{eq.hamiltonian})) so the interaction set is manipulated to achieve this. When determining the existence of an interaction we use Eq.(\ref{eq.traitsum}) as in the procedure used to determine the interaction strengths. But, there is a small but important modification. We can randomly set the parameter ${b_{ij}}\,{\in}\,\{-1,1\}$, and set the second parameter ${c_{ij}}=-{b_{ik}}$ so that a pair of identical phenotypes will incur all modified trait coordinates to be zero. If we set our zeroth autoregression value low in comparison to the interaction threshold then self-interaction is entirely excluded. This effect extends beyond the phenotypic self-interaction to the species description as the correlations ensure that the exclusion occurs, with high probability at least, for phenotypes that are similar but distinct. The autoregression series are created with correlations running from the zero index point in both directions (ie. {0,1,2,...} and {0,99999,99998,...}), thus conforming to the periodic boundary of the trait variables. So, this species level negation of self-interaction is isotropic with regard to a mutation in either direction of a trait.

\section*{\small{Acknowledgements}}

\small{We wish to thank Michel Loreau and Nicolas Loeuille for their valuable perspectives on ecology and providing the impetus for the research theme. A special debt of thanks goes to Andy Thomas and Gunnar Pruessner for continual computational support and provision of highly efficient cluster resources. Simon Laird would like to thank the EPSRC for his PhD funding of this work.}

%\let\oldbibliography\thebibliography
%\renewcommand{\thebibliography}[1]{%
%  \oldbibliography{#1}%
%  \setlength{\itemsep}{0pt}%
%}

%\begin{spacing}{0}
%\bibliographystyle{unsrt}
\bibliographystyle{apacite}
%\bibliographystyle{astron}
%\begin{scriptsize}
\bibliography{paper1}
%\end{scriptsize}
%\end{spacing}
\end{document}